\documentclass[aps,prl,preprint,tightenlines,superscriptaddress,showpacs,byrevtex]{revtex4}

\usepackage{graphicx} 
\usepackage{xspace}
\usepackage{epsfig}
\usepackage{multirow}
\usepackage{dcolumn}  

\newcommand{\de}{\ensuremath{\Delta E}\xspace}

\newcommand{\br}{\ensuremath{\mathcal{B}}\xspace}

\newcommand{\bb}{\ensuremath{B \overline{B}}\xspace}

\def\myspecial#1{}                   
\def\calL{{\mathcal L}}

\def\Mbc{M_{\rm bc}}

\begin{document}

\myspecial{!userdict begin /bop-hook{gsave 300 50 translate 5 rotate
    /Times-Roman findfont 18 scalefont setfont
    0 0 moveto 0.70 setgray
    (\mySpecialText)
    show grestore}def end}




\title{\quad\\[0.5cm] \Large
 Measurements of $B$ Decays to Two Kaons}

\tighten

\affiliation{Budker Institute of Nuclear Physics, Novosibirsk}
\affiliation{Chiba University, Chiba}
\affiliation{Chonnam National University, Kwangju}
\affiliation{University of Cincinnati, Cincinnati, Ohio 45221}
\affiliation{Gyeongsang National University, Chinju}
\affiliation{University of Hawaii, Honolulu, Hawaii 96822}
\affiliation{High Energy Accelerator Research Organization (KEK), Tsukuba}
\affiliation{Hiroshima Institute of Technology, Hiroshima}
\affiliation{Institute of High Energy Physics, Chinese Academy of Sciences, Beijing}
\affiliation{Institute of High Energy Physics, Vienna}
\affiliation{Institute for Theoretical and Experimental Physics, Moscow}
\affiliation{J. Stefan Institute, Ljubljana}
\affiliation{Kanagawa University, Yokohama}
\affiliation{Korea University, Seoul}
\affiliation{Swiss Federal Institute of Technology of Lausanne, EPFL, Lausanne}
\affiliation{University of Ljubljana, Ljubljana}
\affiliation{University of Maribor, Maribor}
\affiliation{University of Melbourne, Victoria}
\affiliation{Nagoya University, Nagoya}
\affiliation{Nara Women's University, Nara}
\affiliation{National Central University, Chung-li}
\affiliation{National United University, Miao Li}
\affiliation{Department of Physics, National Taiwan University, Taipei}
\affiliation{H. Niewodniczanski Institute of Nuclear Physics, Krakow}
\affiliation{Nippon Dental University, Niigata}
\affiliation{Niigata University, Niigata}
\affiliation{Nova Gorica Polytechnic, Nova Gorica}
\affiliation{Osaka City University, Osaka}
\affiliation{Osaka University, Osaka}
\affiliation{Panjab University, Chandigarh}
\affiliation{Peking University, Beijing}
\affiliation{Princeton University, Princeton, New Jersey 08544}
\affiliation{Saga University, Saga}
\affiliation{University of Science and Technology of China, Hefei}
\affiliation{Seoul National University, Seoul}
\affiliation{Shinshu University, Nagano}
\affiliation{Sungkyunkwan University, Suwon}
\affiliation{University of Sydney, Sydney NSW}
\affiliation{Tata Institute of Fundamental Research, Bombay}
\affiliation{Toho University, Funabashi}
\affiliation{Tohoku Gakuin University, Tagajo}
\affiliation{Tohoku University, Sendai}
\affiliation{Department of Physics, University of Tokyo, Tokyo}
\affiliation{Tokyo Institute of Technology, Tokyo}
\affiliation{Tokyo Metropolitan University, Tokyo}
\affiliation{Tokyo University of Agriculture and Technology, Tokyo}
\affiliation{University of Tsukuba, Tsukuba}
\affiliation{Virginia Polytechnic Institute and State University, Blacksburg, Virginia 24061}
\affiliation{Yonsei University, Seoul}
  \author{K.~Abe}\affiliation{High Energy Accelerator Research Organization (KEK), Tsukuba} 
  \author{I.~Adachi}\affiliation{High Energy Accelerator Research Organization (KEK), Tsukuba} 
  \author{H.~Aihara}\affiliation{Department of Physics, University of Tokyo, Tokyo} 
  \author{Y.~Asano}\affiliation{University of Tsukuba, Tsukuba} 
  \author{V.~Aulchenko}\affiliation{Budker Institute of Nuclear Physics, Novosibirsk} 
  \author{T.~Aushev}\affiliation{Institute for Theoretical and Experimental Physics, Moscow} 
  \author{A.~M.~Bakich}\affiliation{University of Sydney, Sydney NSW} 
  \author{V.~Balagura}\affiliation{Institute for Theoretical and Experimental Physics, Moscow} 
  \author{S.~Banerjee}\affiliation{Tata Institute of Fundamental Research, Bombay} 
  \author{E.~Barberio}\affiliation{University of Melbourne, Victoria} 
% ...
\collaboration{The Belle Collaboration}

\begin{abstract}
We report measurements of $B$ meson decays to two kaons  
 using 253 fb$^{-1}$ of data collected with the Belle detector at the
KEKB energy-asymmetric $e^+e^-$ collider.  We find evidence for signals in 
$B^+\to \overline K^0 K^+$ and $B^0\to K^0 \overline{K}^0$ with significances of $3.0
\sigma$ and $3.5 \sigma$, respectively. (Charge-conjugate modes included)
The corresponding branching fractions are measured to be
$\br(B^+\to \overline{K}^0 K^+) = (1.0\pm 0.4 \pm 0.1)\times 10^{-6}$ and 
$\br(B^0\to K^0 \overline{K}^0) = (0.8\pm 0.3 \pm 0.1)\times 10^{-6}$.
These decay modes are examples of hadronic $b \to d$ transitions. No signal is 
observed in the decay $B^0\to K^+ K^-$ and we set an upper limit of 
$3.7 \times 10^{-7}$ at 90\% confidence level.    

\end{abstract}


\maketitle

\tighten

{\renewcommand{\thefootnote}{\fnsymbol{footnote}} 
\setcounter{footnote}{0}




Recent precise measurements of the branching fractions \cite{br} and 
partial rate asymmetries \cite{acp} from the decays $B\to K\pi, \pi\pi$ 
provide essential information to understand the $B$ decay mechanism and
to probe possible contributions from new physics.
The rates for these decays constrain the hadronic $b \to s$ and $b \to u$
amplitudes. Here we report results on $B^0 \to K^0 \overline{K}^0$ and
$B^+ \to \overline{K}^0 K^+$ decays, which are examples of $b \to d$ hadronic
transitions.
We also discuss a search for $B^0\to K^+ K^-$, which is sensitive to effects
of final-state interactions (FSI)~\cite{fsi}.
The results are based on a sample of 275 million $\bb$
pairs collected with the Belle detector at the KEKB $e^+e^-$ asymmetric-energy
(3.5 on 8~GeV) collider~\cite{KEKB} operating at the $\Upsilon(4S)$ resonance.

The Belle detector is a large-solid-angle magnetic
spectrometer that consists of a silicon vertex detector (SVD),
a 50-layer central drift chamber (CDC), an array of
aerogel threshold Cherenkov counters (ACC),
a barrel-like arrangement of time-of-flight
scintillation counters (TOF), and an electromagnetic calorimeter (ECL)
comprised of CsI(Tl) crystals located inside
a superconducting solenoid coil that provides a 1.5~T
magnetic field.  An iron flux-return located outside
the coil is instrumented to detect $K_L^0$ mesons and to identify
muons (KLM).  The detector is described in detail elsewhere~\cite{Belle}.
Two different inner detector configurations were used. For the first sample 
of 152 million $\bb$ pairs (Set I), a 2.0 cm radius beampipe
and a 3-layer silicon vertex detector were used;
for the latter  123 million $\bb$ pairs (Set II),
a 1.5 cm radius beampipe, a 4-layer silicon detector
and a small-cell inner drift chamber were used\cite{Ushiroda}.
   
Charged kaons are required to have a distance of closest approach to the 
interaction point (IP) in the beam direction ($z$) of less than 4 cm and
less than 0.1 cm in  the transverse plane.  
Charged kaons and pions are identified using $dE/dx$
information and Cherenkov light yields in the ACC.
The $dE/dx$ and ACC information are combined to form
a $K$-$\pi$ likelihood ratio, 
$\mathcal{R}(K/\pi) = \mathcal{L}_K/(\mathcal{L}_K+\mathcal{L}_\pi)$,
where $\mathcal{L}_{K}$ $(\mathcal{L}_{\pi})$ is the likelihood that the track
is a kaon (pion).  Charged tracks with $\mathcal{R}(K/\pi)>0.6$ are
regarded as kaons. Furthermore,
charged tracks that are positively identified as electrons or muons are 
rejected.
The electron identification uses information composed of $E/p$ and $dE/dx$,
shower shape, track matching $\chi^2$, and ACC light yields, while
information from the KLM, $dE/dx$ and ACC are combined to identify muons.
The kaon identification efficiency and misidentification
rate are determined from a sample of kinematically identified
$D^{*+}\to D^0\pi^+, D^0\to K^-\pi^+$ decays, where the kaons  from the
$D$ decay are selected in the same kinematic region as in
$B \to K \overline K$ decays. The kaon efficiency is measured to be 
$(84.24\pm 0.13)\%$ for Set I and $(82.84\pm 0.14)\%$ for Set II, while the 
pion-fake-kaon rates are $(5.40\pm 0.08)\%$ and $(6.86\pm 0.11)\%$,
respectively.
  
Candidate $K^0$ mesons are reconstructed through the $K_S^0 \to \pi^+\pi^-$ 
decay. We pair oppositely-charged tracks assuming the pion hypothesis
and require the invariant mass of the 
pair to be within 18 MeV/$c^2$ of the nominal $K_S^0$ mass. Furthermore, the 
intersection point of the $\pi^+ \pi^-$ pair must be displaced from the IP.

Two variables are used to identify $B$ candidates: the beam-constrained mass,
$M_{\rm bc} \equiv  
\sqrt{E^{*2}_{\mbox{\scriptsize beam}} - p_B^{*2}}$, and the energy difference,
$\Delta E \equiv E_B^* - E^*_{\mbox{\scriptsize beam}}$, where 
$E^*_{\mbox{\scriptsize beam}}$ is the run dependent beam energy and 
$E^*_B$ and $p^*_B$ are
the reconstructed energy and momentum of the $B$ candidates in the
center-of-mass (CM) frame, respectively. Events with 
$M_{\rm bc} > 5.20$ GeV/$c^2$ and $|\Delta E| < 0.3~{\rm GeV}$
are selected for analysis. 

The dominant background is from $e^+e^- \to q\bar q ~( q=u,d,s,c )$ continuum
events. Event topology and $B$ flavor tagging information are used to
distinguish between the spherically distributed $B\overline{B}$ events and 
the jet-like continuum backgrounds.
We combine a set of modified Fox-Wolfram moments \cite{pi0pi0} into a
Fisher discriminant. A signal/background likelihood is formed, based on a 
GEANT-based~\cite{geant}
Monte Carlo (MC) simulation, from the product of the probability density 
function (PDF) for the Fisher discriminant and that for the cosine of the angle
between the $B$ flight direction and the positron beam. The continuum
suppression is achieved by applying a requirement on a likelihood ratio
$\mathcal{R} = {\calL}_s/({\calL}_s + {\calL}_{q \bar{q}})$, where
${\calL}_{s (q \bar{q})}$ is the signal ($q \bar{q}$) likelihood. Additional
background discrimination is provided by $B$ flavor tagging. 
For each event, the standard Belle flavor tagging algorithm \cite{tagging} 
provides a discrete variable indicating the probable flavor of the tagging $B$
meson, and a quality $r$, a continuous variable ranging from zero for no
flavor tagging information to unity for unambiguous flavor assignment.
An event with a high value of $r$ (typically
containing a high-momentum lepton) is more likely to be a $B \overline B$ 
event, and a looser $\mathcal{R}$ requirement can be applied. We divide the
data into $r>0.5$ and $r \le 0.5$ regions. A selection requirement on 
 $\mathcal{R}$ for events in each $r$ region of Set I and Set II is applied 
according to a figure-of-merit defined as
$N_s^{\rm exp}/\sqrt{N_s^{\rm exp}+N_{q\bar{q}}^{\rm exp}}$, where
$N_s^{\rm exp}$ denotes the expected signal yields based on MC simulation
and the assumed branching fractions, $1.0 \times 10^{-6}$,  and
$N_{q\bar{q}}^{\rm exp}$ denotes the
expected $q\bar q$ yields from sideband data ($M_{\rm bc}<5.26$ GeV/$c^2$).

Background contributions from $\Upsilon(4S) \to B\overline B$ events are 
investigated using a large MC sample, which includes events from $b\to c$
transitions and charmless decays. After all the selection requirements,
no $B\overline{B}$ background is found for the $B^0 \to K^0 \overline K^0$
mode. Owing to $K$-$\pi$ misidentification, large $B^0\to K^+\pi^-$ and
$B^+\to K^0 \pi^+$ feed-across backgrounds appear in the $B^0\to K^+ K^-$
and $B^+\to \overline{K}^0 K^+$ modes, 
respectively. A small charmless three-body contribution is found at 
low $\Delta E$ values for these two modes.

The signal yields are extracted by performing unbinned two dimensional
maximum likelihood (ML) fits to the ($M_{\rm bc}$, $\Delta E$)
distributions.
The likelihood for each mode is defined as
\begin{eqnarray*}
\mathcal{L} & = & {\rm exp}\; (-\sum_{s,k,j} N_{s,k,j}) 
\prod_i (\sum_{s,k,j} N_{s,k,j} {\mathcal P}_{s,k,j,i}) \;\;\; 
\\
\mathcal{P}_{s,k,j,i} & = & 
P_{s,k,j}(M_{{\rm bc}\;i}, \Delta E_i),  
\end{eqnarray*}
where $s$ indicates Set I or Set II, $k$ distinguishes between events
in the $r<0.5$ and $r>0.5$ regions, $i$ is the identifier of
the $i$-th event, $P(M_{\rm bc}, \Delta E)$ is the two-dimensional PDF of
$M_{\rm bc}$ and $\Delta E$, 
$N_j$ is the number of events for the
category $j$, which corresponds to either signal, $q\bar{q}$ continuum,
a feed-across due to $K$-$\pi$ misidentification, or
background from other charmless three-body $B$ decays.  

All the signal PDFs ($P_{s,k,j=\mathrm{signal}}(M_{\rm bc},\Delta E)$)
are parametrized by a product of a 
single Gaussian for $M_{\rm bc}$ and a double Gaussian for $\Delta E$    
using MC simulations based on the Set I and Set II detector configurations. 
The same signal PDFs are used for events in the two different $r$ 
regions. Since the $M_{\rm bc}$ signal distribution is dominated by
the beam energy spread, we use the signal peak positions and resolutions
obtained from $B^+ \to \overline{D}{}^0\pi^+$ data
($\overline{D}{}^0 \to K^0_S\pi^+ \pi^-$ sub-decay is used for the $K^0 \overline K^0$
mode, while $\overline{D}{}^0\to K^+\pi^-$ is used for the other two modes) with
small mode dependent correlations obtained from MC.
The MC-predicted $\Delta E$ resolutions are verified using the
invariant mass distributions of high momentum $D$ mesons. The decay mode 
$\overline{D}{}^0\to K^+\pi^-$ is used for $B^0\to K^+ K^-$, 
$D^+\to K^0_S \pi^+$ for $B^+\to K^0\pi^+$ and  
$\overline{D}{}^0\to K^0_S\pi^+\pi^-$ for $B^0\to K^0 \overline K^0$.
The parameters  that describe the shapes of the PDFs are fixed in all
of the fits.

The continuum background in $\de$ is described by a linear function
while the $\Mbc$ distribution is parameterized by an
ARGUS function $f(x) = x \sqrt{1-x^2}\;{\rm exp}\;[ -\xi (1-x^2)]$, where
$x$ is $\Mbc$ divided by half of the total center of mass energy \cite{argus}.
Therefore, the continuum PDF is the product of this ARGUS function and the
linear function, where the overall normalization, $\xi$ and the slope of the
linear function are free parameters in the fit. 
These free parameters are $r$-dependent and allowed to be different in Set I 
and Set II.  The background PDFs for charmless three-body $B$ decays for the 
$K^+ K^-$ and $\overline{K}^0K^+$ modes are each modeled by a smoothed two-dimensional 
histogram, obtained from a large MC sample. The feed-across backgrounds for
these two modes from  the $K^+\pi^-$ and $K^0\pi^+$ events have
$\Mbc - \de$ shapes similar to the signals with the $\de$ peak positions
shifted by $\simeq 45$ MeV. The methods to model the $K^+K^-$ and
$\overline{K}^0 K^+$ signal PDFs are also applied to describe the feed-across background.

\begin{figure*}[htb]
\hspace{-1.0cm}
\includegraphics[width=0.95\textwidth]{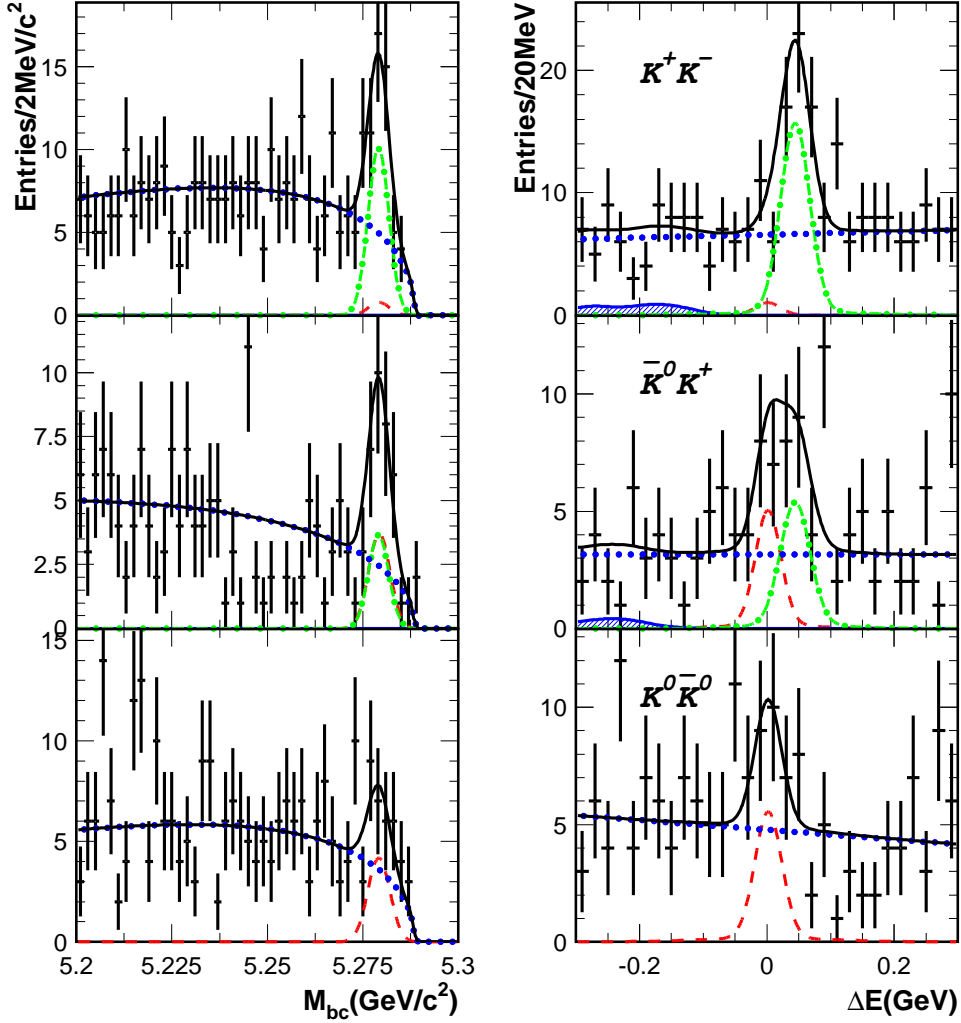}
\caption{$M_{\rm bc}$ (left) and $\Delta E$ (right) distributions for
$B^0\to K^+K^-$ (top) and $B^+\to \overline{K}^0 K^+$ (middle) and
$B^0\to K^0 \overline K^0$ candidates. The histograms show
the data, while the curves represent the various components from
the fit: signal (dashed), continuum (dotted), three-body $B$ decays
(hatched), background from  mis-identification (dash-dotted),
and sum of all components (solid). In the $K^+ K^-$ mode, there is 
a large contribution from misidentified $K^+ \pi^-$ but no significant
signal excess. In the $\overline{K}^0 K^+$ mode, the signal and misidentified
$K^0 \pi^+$ contributions are comparable in size. In the $K^0 \overline{K}^0$
mode, there is a signal excess but no misidentification background.}
\label{fig:kk}
\end{figure*}

When likelihood fits are performed, the yield for each background
component ($N_{s,k,j}$ where
$j=q\bar{q}$, feed-across, charmless) is allowed to
float independently for each $s$ (Set I or Set II), and $k$ bin (low
or high $r$ region). For the signal component, the same branching fraction
is required by constraining the number of signal events in each $(s,k)$
bin using the measured efficiency in the corresponding $(s,k)$ bin.
Table \ref{tab:kk} summarizes the fit results  for each mode. We observe
$13.3 \pm 5.6\pm 0.6$ ${K}^0 K^+$ and $15.6 \pm 5.8^{+1.1}_{-0.6}$
$K^0 \overline K^0$ signal events with 
significances of $3.0 \sigma$ and $3.5 \sigma$, respectively. The second 
errors in the yields are the systematic errors from fitting, estimated
from the deviations after varying each parameter of the
signal PDFs by one standard deviation, and from modeling
the three-body background, studied by excluding the low $\Delta E$ region
($<-0.15$ GeV) and repeating the fit. At each step, the yield deviation is 
added in quadrature to provide the fitting systematic errors and the  
statistical significance is computed by taking the square root of the 
difference between the value of  $-2\ln\mathcal L$  for the best fit value and 
zero signal yield. The 
smallest value is chosen to be the significance including the systematic 
uncertainty.       

Figure ~\ref{fig:kk} shows the $\Mbc$ and $\de$ projections of the fits 
after requiring events to have $|\de|<0.06$ GeV and $5.271$
GeV/$c^2 < {M_{\rm bc}} < 5.289$
GeV/$c^2$, respectively. The feed-across yields are 
$47.1\pm 8.7$  in the $K^+K^-$ mode and $16.4 \pm 6.1$  
in the ${K}^0 K^+$ mode. The amounts of the feed-across background are 
consistent with the expectations of
49.1 $K^+\pi^-$ and 18.8 $K^0\pi^+$ events, based on MC 
simulation and measured branching fractions \cite{hfag}.       
The MC modeling of the requirement on the likelihood ratio, $\mathcal{R}$
is investigated 
using the  $B^+\to \overline{D}{}^0\pi^+ (\overline{D}{}^0\to K^0_S\pi^+\pi^-$ 
for $K^0K^0$ and $\overline{D}{}^0\to K^+\pi^-$ for the others)  
samples. The obtained systematic errors are $\pm 2.9\%$ for
$B^0\to K^0 \overline K^0$ and $\pm 6.8\%$  for the other two modes.  The
systematic error on the charged track reconstruction efficiency is
estimated to be around $1$\% per track using partially
reconstructed $D^*$ events.
The resulting $K_S^0$ reconstruction is verified by comparing the
ratio of $D^+\to K_S^0\pi^+$ and $D^+\to K^-\pi^+\pi^+$ yields 
with the MC expectation. The resulting $K_S^0$ detection systematic 
error is $\pm4.5\%$.  
The final systematic errors are then obtained by quadratically summing the
errors due to the reconstruction efficiency and the fitting systematics.

\begin{table*}[htb]
\begin{center}
\caption{Fitted signal yields, reconstruction efficiencies, product of
 efficiencies and sub-decay branching fractions $({\cal B}_s)$, branching
 fractions and significances for individual modes.}
\begin{tabular}{lccccc}
\hline\hline
~Mode~ & Yield & Eff.(\%) & Eff.$\times {\cal B}_s$ (\%)& ${\cal B}(10^{-6})$& Sig.\\
\hline
~$K^+K^-$ &$2.5^{+5.1}_{-4.1}$ & $15.5$& $15.5$ & $<0.37$& 0.5\\
~$\overline{K}^0K^+$ &$13.3\pm 5.6$  &$14.5$ & $5.0$ &$1.0\pm 0.4\pm 0.1$& 3.0\\
~$K^0 \overline K^0$ &$15.6\pm 5.8$ &$28.7$  & $6.8$ &$0.8\pm 0.3\pm 0.1$ & 3.5\\
\hline\hline
\end{tabular}
\label{tab:kk}
\end{center}
\end{table*}

With 275 million $B\overline{B}$ pairs, we find evidence of $B^+\to \overline{K}^0 K^+$ 
and $B^0\to K^0 \overline K^0$ with branching fractions
$\br(B^+\to \overline{K}^0 K^+) = (1.0\pm 0.4 \pm 0.1)\times 10^{-6}$ and
$\br(B^0\to K^0 \overline K^0) = (0.8\pm 0.3 \pm 0.1)\times 10^{-6}$. These are
examples of hadronic $b \to d$ transitions. Our measurements 
are consistent with preliminary results reported by the BaBar collaboration
and agree with some theoretical predictions
\cite{bjorken, pqcd, pqcd_kk, fleischer, chiang}. It has been suggested that
the branching fraction and CP asymmetry of the mode $B^0 \to K^0 \overline{K}^0$,
which originates from the flavor-changing neutral current process
$\bar{b} \to \bar{d} s \bar{s}$, may be sensitive to physics beyond the
Standard Model \cite{fleischer}. Measurements with larger statistics are
needed for this purpose. No signal is observed in $B^0\to
K^+K^-$ and we set the upper limit of $3.7 \times 10^{-7}$ at the 90\%
confidence level, using the Feldman-Cousins approach \cite{feldman} taking
into account both the statistical and systematic errors \cite{pole}.

We thank the KEKB group for the excellent operation of the
accelerator, the KEK cryogenics group for the efficient
operation of the solenoid, and the KEK computer group and
the NII for valuable computing and Super-SINET network
support.  We acknowledge support from MEXT and JSPS (Japan);
ARC and DEST (Australia); NSFC (contract No.~10175071,
China); DST (India); the BK21 program of MOEHRD and the CHEP
SRC program of KOSEF (Korea); KBN (contract No.~2P03B 01324,
Poland); MIST (Russia); MHEST (Slovenia);  SNSF (Switzerland); NSC and MOE
(Taiwan); and DOE (USA).


\end{document}